\pdfoutput=1

\documentclass[11pt,twoside,a4paper,cmspaper,final,collab]{cms-tdr}
\def\svnVersion{333043}\def\cmsCernNoTag{CERN-PH-EP/2013-037}\def\cmsCernDate{\today}\def\cmsMessage{Published in Physics Letters B as \href{http://dx.doi.org/10.1016/j.physletb.2016.02.047}{\doi{10.1016/j.physletb.2016.02.047.}}}

\begin{document}\cmsNoteHeader{BPH-14-002}

\hyphenation{had-ron-i-za-tion}
\hyphenation{cal-or-i-me-ter}
\hyphenation{de-vices}
\RCS$HeadURL: svn+ssh://svn.cern.ch/reps/tdr2/papers/BPH-14-002/trunk/BPH-14-002.tex $
\RCS$Id: BPH-14-002.tex 330524 2016-03-04 01:52:06Z eduard $
\newlength\cmsFigWidth
\ifthenelse{\boolean{cms@external}}{\setlength\cmsFigWidth{0.98\columnwidth}}{\setlength\cmsFigWidth{0.7\textwidth}}
\ifthenelse{\boolean{cms@external}}{\providecommand{\cmsLeft}{top}}{\providecommand{\cmsLeft}{left}}
\ifthenelse{\boolean{cms@external}}{\providecommand{\cmsRight}{bottom}}{\providecommand{\cmsRight}{right}}
\providecommand{\cPKK}{\ensuremath{\mathrm{K}^+\mathrm{K}^-}\xspace}
\providecommand{\cPpipi}{\ensuremath{\pi^+\pi^{-}}\xspace}
\providecommand{\cPfz}{\ensuremath{\mathrm{f}_0}\xspace}
\renewcommand{\Pfz}{\ensuremath{\cPfz(980)}\xspace}
\providecommand{\JPsiX}{\ensuremath{\JPsi\,}}
\cmsNoteHeader{BPH-14-002}
\title{Measurement of the ratio $\mathcal{B}( \PBzs \to\JPsiX \cPfz(980))/\mathcal{B}(\PBzs\to\JPsiX \phi(1020))$ in pp
collisions at $\sqrt{s}=7\TeV$}

\author{The CMS Collaboration}

\date{\today}

\abstract{
A measurement of the ratio of the branching fractions of the $\PBzs$ meson to
$\JPsiX\Pfz$ and to $\JPsiX \phi(1020)$ is presented. 
The $\JPsi$, $\Pfz$, and $\phi(1020)$ are observed through their decays to
$\mu^+\mu^-$, $\cPpipi$, and $\cPKK$, respectively.  
The $\cPfz$ and the $\phi$ are identified by requiring $\abs{M_{\cPpipi}-974\MeV}<50\MeV$ and $\abs{M_{\cPKK}-1020\MeV}<10\MeV$.
The analysis is based on a data sample of pp collisions
at a centre-of-mass energy of $7\TeV$, collected by the CMS experiment at the LHC,
corresponding to an integrated luminosity of 5.3\fbinv.
The measured ratio is
$\frac{\mathcal{B}(\PBzs\to  \JPsiX\cPfz)\,\mathcal{B}
(\cPfz\to \Pgpp\Pgpm)}{\mathcal{B}(\PBzs\to  \JPsiX\phi )\,\mathcal{B}(\phi\to \PKp\PKm)}
= 0.140 \pm  0.008\stat \pm 0.023\syst$,
where the first uncertainty is statistical and the second is systematic.
}

\hypersetup{%
pdfauthor={CMS Collaboration},%
pdftitle={Measurement of the ratio B(B[0,s] to J/psi f0(980) )/B(B[0,s] to J/psi phi(1020)) in pp collisions at sqrt(s) = 7 TeV},%
pdfsubject={CMS},%
pdfkeywords={CMS, B Physics}}

\maketitle

\section{Introduction}
Since the observation of the decay $\PBzs \to \JPsiX\pi^+\pi^-$ with $\JPsi \to \Pgmp\Pgmm$,
and the $\pi^+\pi^-$ mass spectrum indicating a large $\Pfz$ component \cite{LHCbBRf0:2011},
this channel has been regarded with great interest in heavy-flavor physics.
More detailed studies of the $\pi^+\pi^-$ mass spectrum have shown the $\pi^+\pi^-$ system to be
almost entirely CP odd \cite{JpsiPiPiLHCb:2012, LHCbCPcomp:2014}.
This opens up the possibility of directly measuring the lifetime of the CP-odd part of the $\PBzs$
meson \cite{CDFBRf0:2011, LHCbLTf0:2012}.
In addition, the $\PBzs \to \JPsiX\pi^+\pi^-$ decay
has been used for the measurement of the CP-violating
phase $\phi_s$ \cite{LHCbPhif0:2012,LHCbPhif0:2014},
making an important contribution to the world-average value of $\phi_s$
\cite{JpsiPhiD0:2008,  JpsiPhiCDF:2008, MaganaD0:2009, LHCbPhi_s:2012, AtlasPhi_s:2012,CMSphis}.
The phase $\phi_{s}$ is predicted to be small in the standard model \cite{Charles:2011}, making its determination
interesting because of the large enhancements that can be introduced by new physics
\cite{Ball/Fleischer:2006,Lenz:2007}. In what follows, we will refer to the $\Pfz$ as
$\cPfz$ and the $\Pgf$ as $\phi$.

This Letter presents the measurement of the ratio $R_{\cPfz/\phi}$ of the
branching fractions $\mathcal{B}( \PBzs \to  \JPsiX\cPfz )\mathcal{B}(\cPfz\to \Pgpp\Pgpm)$
and $\mathcal{B}( \PBzs \to  \JPsiX \phi )\mathcal{B}(\phi\to \PKp\PKm)$, where in both cases the $\JPsi$ is
detected through its decay to $\Pgmp\Pgmm$. The $\cPfz$ and the $\phi$
are identified by requiring $\abs{M_{\cPpipi}-974\MeV}<50\MeV$ and $\abs{M_{\cPKK}-1020\MeV}<10\MeV$.
The appearance of $\PBzs \to  \JPsiX\cPfz $ decays was first discussed in \cite{Stone/Zhang:2009} with a theoretical estimate for $R_{\cPfz/\phi}$
of approximately $0.2$, which is consistent with results from several 
experiments \cite{JpsiPiPiLHCb:2012,CDFBRf0:2011, BelleBRf0:2011,D0BRf0:2012}.
Detailed studies of the $\pi^+\pi^-$ mass spectrum of the $\PBzs \to \JPsiX\pi^+\pi^-$ decay in 
$0.3<M_{\cPpipi}<2.5\GeV$ \cite{JpsiPiPiLHCb:2012, LHCbCPcomp:2014}
reveal this final state to have contributions from several resonances in $M_{\cPpipi}$, and the $\cPfz$ component to range from 65.0\% to 94.5\%. However,
according to the same results, the contaminations from other resonances in $\abs{M_{\cPpipi} - 974\MeV} < 50\MeV$ are several orders of magnitude lower
than the $\cPfz$ component, including the non-resonant S-wave. Based on this, the measurement of $R_{\cPfz/\phi}$ is performed assuming that
the selected region of $M_{\cPpipi}$ is dominated by $\PBzs \to  \JPsiX \cPfz$ decays and neglecting other resonances.
Systematic uncertainties are assigned to the measurement owing to these assumptions, taking into account the uncertainty 
in the $\cPfz$ component and the interferences with other resonances in the selected mass window for $M_{\cPpipi}$.

Experimentally, the ratio $R_{\cPfz/\phi}$ is given by
\begin{linenomath}
\begin{equation}\label{eq:BR}
R_{\cPfz/\phi} =  \frac{\mathcal{B}(\PBzs \to  \JPsiX \cPfz)\,\mathcal{B}(\cPfz\to \Pgpp\Pgpm)}{\mathcal{B}
(\PBzs \to  \JPsiX \phi)\,\mathcal{B}(\phi\to \PKp\PKm)} = \frac{N_{\text{obs}}^{\cPfz}}
{N_{\text{obs}}^{\phi}} \,
\epsilon_{\text{reco}}^{\phi/\cPfz},
\end{equation}
\end{linenomath}
where $N_{\text{obs}}^{\cPfz}$ and $N_{\text{obs}}^{\phi}$ are the observed yields of $\PBzs \to
\JPsi(\Pgmp\Pgmm)\cPfz$  with  $\cPfz\to \Pgpp\Pgpm$ and $\PBzs \to \JPsi(\Pgmp\Pgmm) \phi$  with
$\phi\to \PKp\PKm$ decays, respectively, and $\epsilon^{\phi/\cPfz}_{\text{reco}}$
is the ratio of the detection efficiencies for the $\PBzs$ decay mode with a $\phi$ to the decay mode with a $\cPfz$.
Uncertainties in the b quark production cross section cancel in the ratio, as do those from the $\JPsi \to \MM$ branching
fraction and the integrated luminosity.  Given the similar topologies of the two final states, systematic uncertainties
related to the tracking efficiency and the muon identification also cancel in the ratio.

\section{The CMS detector}

The central feature of the CMS apparatus is a superconducting solenoid of 6\unit{m} internal diameter. Within the
3.8\unit{T} field volume are a silicon pixel and strip tracker, a lead tungstate crystal electromagnetic
calorimeter, and a brass and scintillator hadron calorimeter, each composed of a barrel and two endcap sections.
Muons are measured in the pseudorapidity range $\abs{\eta}<2.4$ in gas-ionization detectors embedded in the steel flux-return
yoke outside the solenoid, which are made using three technologies: drift tubes, cathode strip chambers, and
resistive-plate chambers. Extensive forward calorimetry
complements the coverage provided by the barrel and endcap detectors. The main
subdetectors used in this analysis are the silicon tracker and the muon systems.

The silicon tracker measures charged particles within the pseudorapidity range $\abs{\eta}< 2.5$ and consists of
1440 silicon pixel and 15\,148 silicon strip detector modules.
Matching muons to tracks measured in the silicon tracker results
in a relative transverse momentum resolution for muons with $20
<\pt < 100\GeV$ of 1.3--2.0\% in the barrel and better than 6\%
in the endcaps, The \pt resolution in the barrel is better than
10\% for muons with \pt up to 1\TeV~\cite{Chatrchyan:2012xi}.

The first level of the CMS trigger system, composed of custom hardware processors, uses information from the
calorimeters and muon detectors to select the most interesting events in a fixed time interval of less than 4\mus.
The high-level trigger (HLT) processor farm further decreases the event rate
to less than 1\unit{kHz}, before data storage.

A more detailed description of the CMS detector, together with a definition of the coordinate system used and the
relevant kinematic variables, can be found in Ref. \cite{Chatrchyan:2008zzk}.

\section{Event selection}

The data sample used for this measurement was collected in 2011 by the CMS experiment at the CERN LHC in proton-proton
collisions at a centre-of-mass energy of 7\TeV and corresponds
to an integrated luminosity of 5.3\fbinv.

The search for $\PBzs \to \JPsiX \cPfz$ decays is performed in events with two muon candidates
selected by the dimuon trigger at the HLT, requiring the muon pair to originate from a displaced vertex. The
dimuon candidates are further required to comply with $\mathrm{L}_{xy}/\sigma_{xy}>3$, where $\mathrm{L}_{xy}$ is the magnitude of the vector $\vec{\mathrm{L}}_{xy}$, which lies in a plane transverse to the beam axis and points from the interaction point
to the dimuon
vertex, and $\sigma_{xy}$ is its uncertainty;
$\cos \alpha_{\JPsi}>0.9$, where $\alpha_{\JPsi}$ is the angle between the direction of the dimuon transverse momentum and
$\vec{\mathrm{L}}_{xy}$;
$\pt > 4\GeV$ and $\abs{\eta}<2.2$ for each muon candidate;
$\pt > 7\GeV$ for the dimuon;
the distance of closest approach of each muon track
with respect to the other muon track $<$0.5\cm.

Reconstruction of the $\PBzs \to \JPsiX\cPfz$ decays begins with the search for $\JPsi$
candidates by combining two muons
of opposite charge to form a vertex with a fit probability $>$0.5\% and an invariant mass ($M_{\JPsi}$)
within $\abs{M_{\JPsi} - 3097.6\MeV}<150\MeV$.
To search for $\cPfz$ candidates, two tracks of opposite charge assumed to be pions are constrained
to a vertex with a probability $>$5\%.
One pion candidate must have $\pt>1\GeV$ and the other $\pt>2.5\GeV$.
In addition, the $\cPfz$ candidate must have $\pt> 3.5\GeV$ and
$M_{\cPfz}$ in the range $\abs{M_{\cPfz}  - 974\MeV}<50\MeV$.  The 974\MeV is the measured mass of $\cPfz$ signal in data modeled by a
Breit--Wigner function.
This value is consistent with the $\cPfz$ mass from the Particle Data Group \cite{Beringer:1900zz} and the LHCb measurement \cite{LHCbBRf0:2011}.
Finally, a vertex is formed with the $\JPsi$ and $\cPfz$ candidates, constraining the dimuon mass
to the nominal $\JPsi$ mass \cite{Beringer:1900zz}.
The $\PBzs \to \JPsiX \cPfz$ candidates are required to have a vertex probability $>$10\%,
$\pt>13\GeV$,
$\cos \alpha_{\PBzs} >0.994$, where $\alpha_{\PBzs}$ is the angle between the direction of the $\PBzs$ transverse momentum and
the vector
$\vec{\mathrm{L}}_{xy}$,
and a proper decay length $>$100\mum.
The proper decay length is
defined as
$(\vec{\mathrm{L}}_{xy}\cdot \ptvec\, M_\PB/\pt^2)$, where $\ptvec$ is the
transverse
momentum of the
$\PBzs$ candidate and $M_\PB$ is the world-average $\PBzs$ mass
\cite{Beringer:1900zz}. In the case of multiple
$\PBzs$ candidates per event, the one with smallest $\PBzs$ vertex fit $\chi^2$ is selected.
The selection criteria
for the $\PBzs$ candidates are established by maximizing $S/\sqrt{S+B}$, where $S$ is the signal yield
obtained
from Monte Carlo (MC) simulation and $B$ is the background yield taken from sideband regions,
defined as the number of events with a $\Pgmp\Pgmm \Pgpp\Pgpm$ invariant mass in the range $5.27$ to $5.30\GeV$ or $5.43$ to $5.46\GeV$.

The same procedure and selection criteria are applied to the reconstruction of the normalization channel
$\PBzs\to \JPsiX \phi$, except that the invariant mass requirement
$\abs{M_{\phi} - 1020\MeV}<10\MeV$ is tighter than that for the $\cPfz$.

\section{Results}

\begin{figure}[hbt]
\centering
\includegraphics[width=\cmsFigWidth]{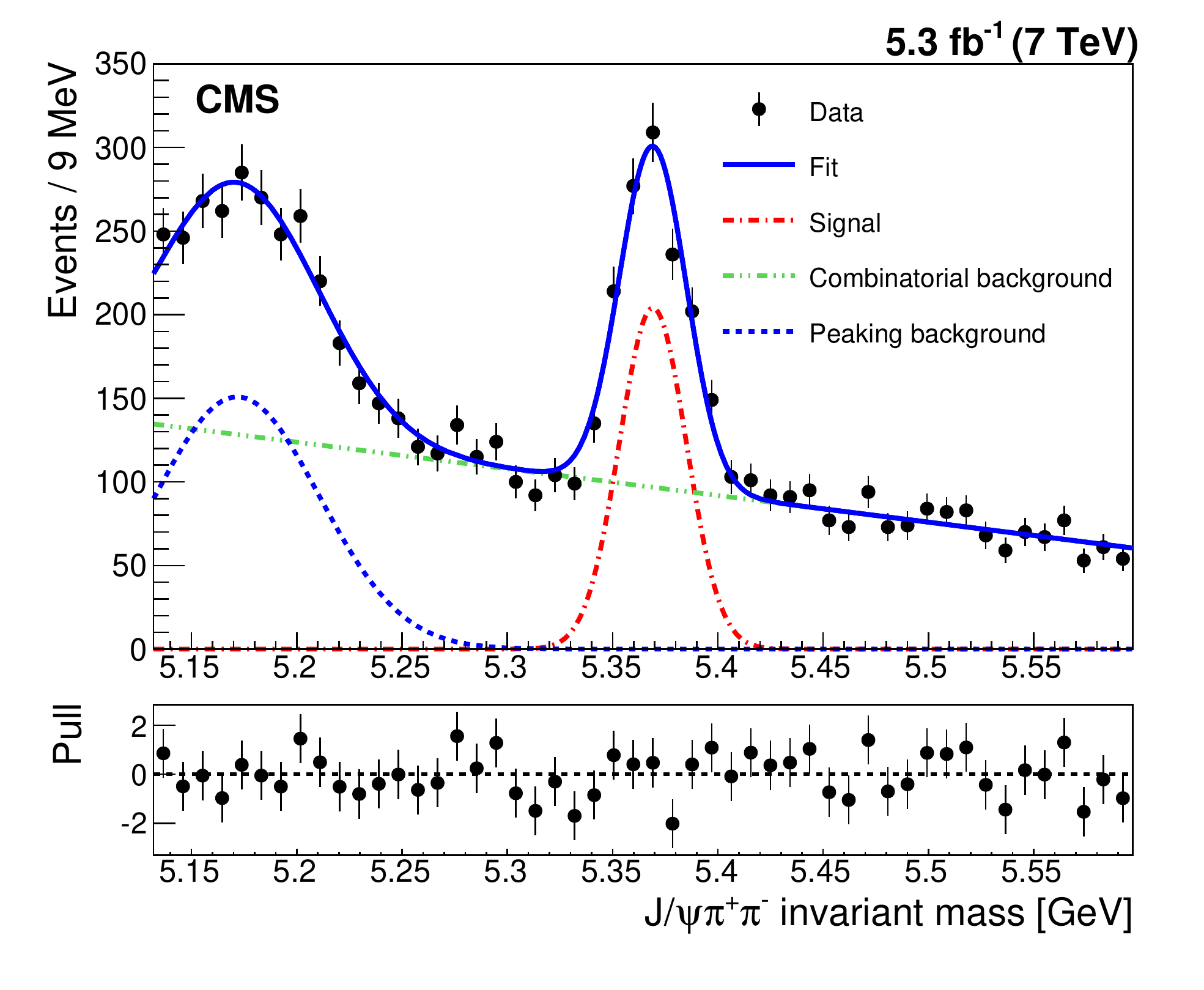}
\caption{Invariant mass distribution of the $\PBzs\to \JPsi (\Pgmp\Pgmm) \cPfz(\Pgpp\Pgpm)$ candidates
(filled circles). The signal is modelled as a Gaussian (dot-dashed line), the combinatorial background  as a first-order
polynomial function
(dashed-double-dotted line), and the peaking background by a Gaussian (dotted line). The result of the total fit
is shown with the solid line. The bottom plot shows the pull, which is the deviation of the data from the fit divided by the uncertainty in the data.}
\label{fig:Bmass}
\end{figure}

The signal yields of both decay channels are extracted using unbinned maximum-likelihood fits of the mass
distributions. The invariant mass distribution of the $\JPsi (\Pgmp\Pgmm)\cPfz(\Pgpp\Pgpm)$ candidates is shown in
Fig.~\ref{fig:Bmass}. It is fit
with a superposition of a Gaussian function representing the signal, a polynomial function to account for the
combinatorial background, and another Gaussian function for any possible peaking background. The latter
models resonant structures that could appear in the left sideband of the $\JPsi (\Pgmp\Pgmm) \cPfz(\Pgpp\Pgpm)$
signal mass owing to the misidentification of a kaon as a pion
coming from decays such as $\PBz\to \JPsiX\PKst(\PKp\Pgpm)$ and
$\PBzs\to \JPsiX \PKp\PKm$, as examples.
In addition, $\PBp\to \JPsiX\PKp(\Pgpp)$ decays can be a source of background when
combined with an extra background pion candidate.
When allowing all parameters to float, the fit returns $N_{\text{obs}}^{\cPfz} = 873 \pm 49$ events and a $\PBzs$ mass of $5369.1
\pm 0.9\MeV$, with a resolution
of $15.9\pm 0.9\MeV$, where the uncertainties are statistical only. The measured values of the $\PBzs$
mass and its resolution are consistent with the MC simulation.

\begin{figure}[hbt]
\centering
\includegraphics[width=\cmsFigWidth]{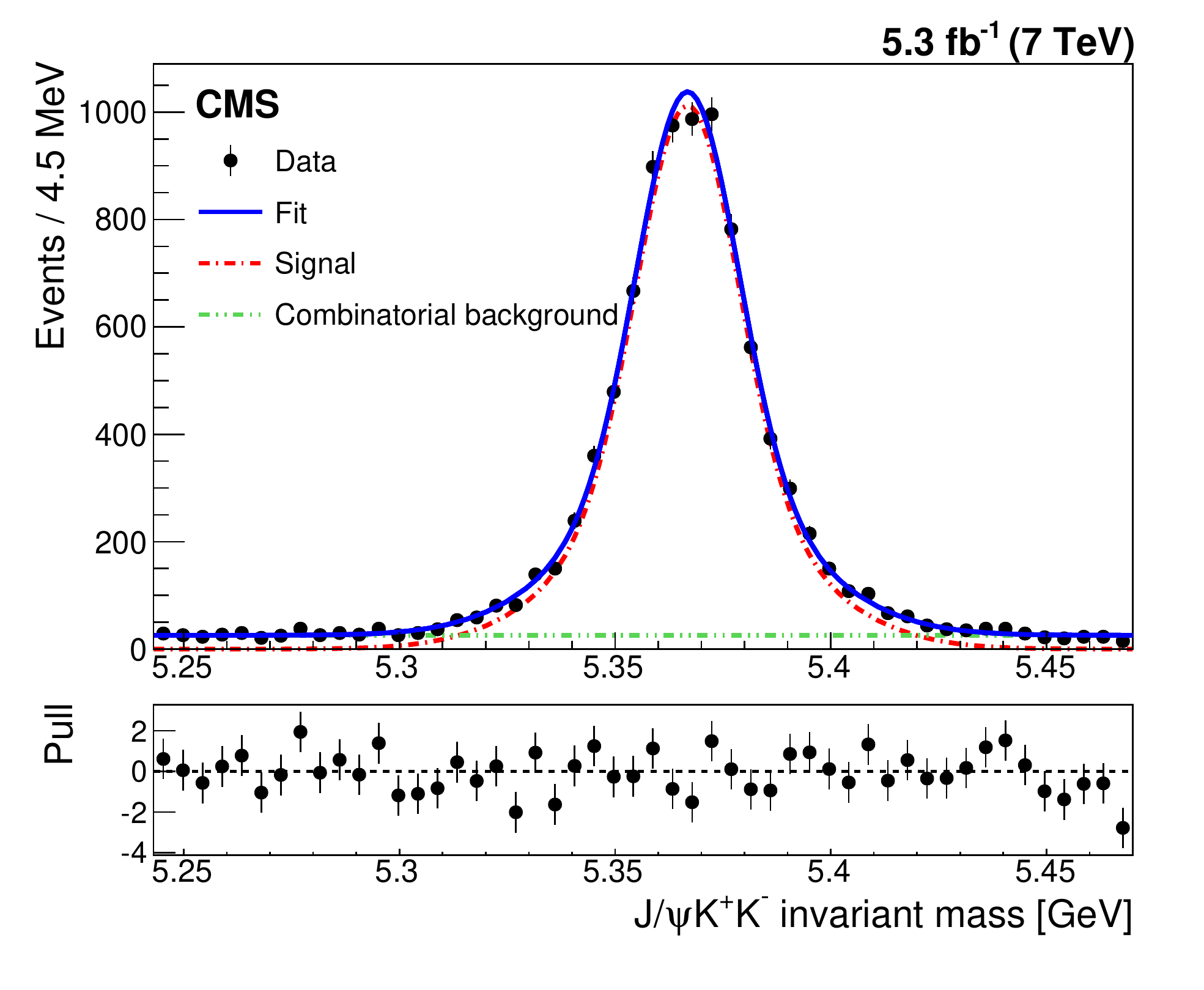}
\caption{Invariant mass distribution of the $\PBzs\to\JPsi(\Pgmp\Pgmm) \phi(\PKp\PKm)$ candidates (black filled circles). The signal model is a double Gaussian (dot-dashed line), while the combinatorial background model
is a constant function (dash-double-dotted line). The total fit is represented by the solid line. The bottom plot shows the deviation of the data to the
fit divided by the statistical uncertainty in the data.}
\label{fig:BmassPhi}
\end{figure}

The $\JPsi(\Pgmp\Pgmm) \phi(\PKp\PKm)$ invariant mass distribution is modelled by two Gaussian functions
for the signal and a constant function for the combinatorial background. A signal yield of
$N_{\text{obs}}^{\phi} = 8377 \pm 107$ events is obtained, with a $\PBzs$ mass of $5366.8\pm 0.2\MeV$
and a resolution of $17.1 \pm 0.1\MeV$, which are consistent with the MC simulation. The corresponding invariant mass distribution is
presented in Fig. \ref{fig:BmassPhi}.

\begin{figure}[hbt]
\centering
\includegraphics[width=0.49\textwidth]{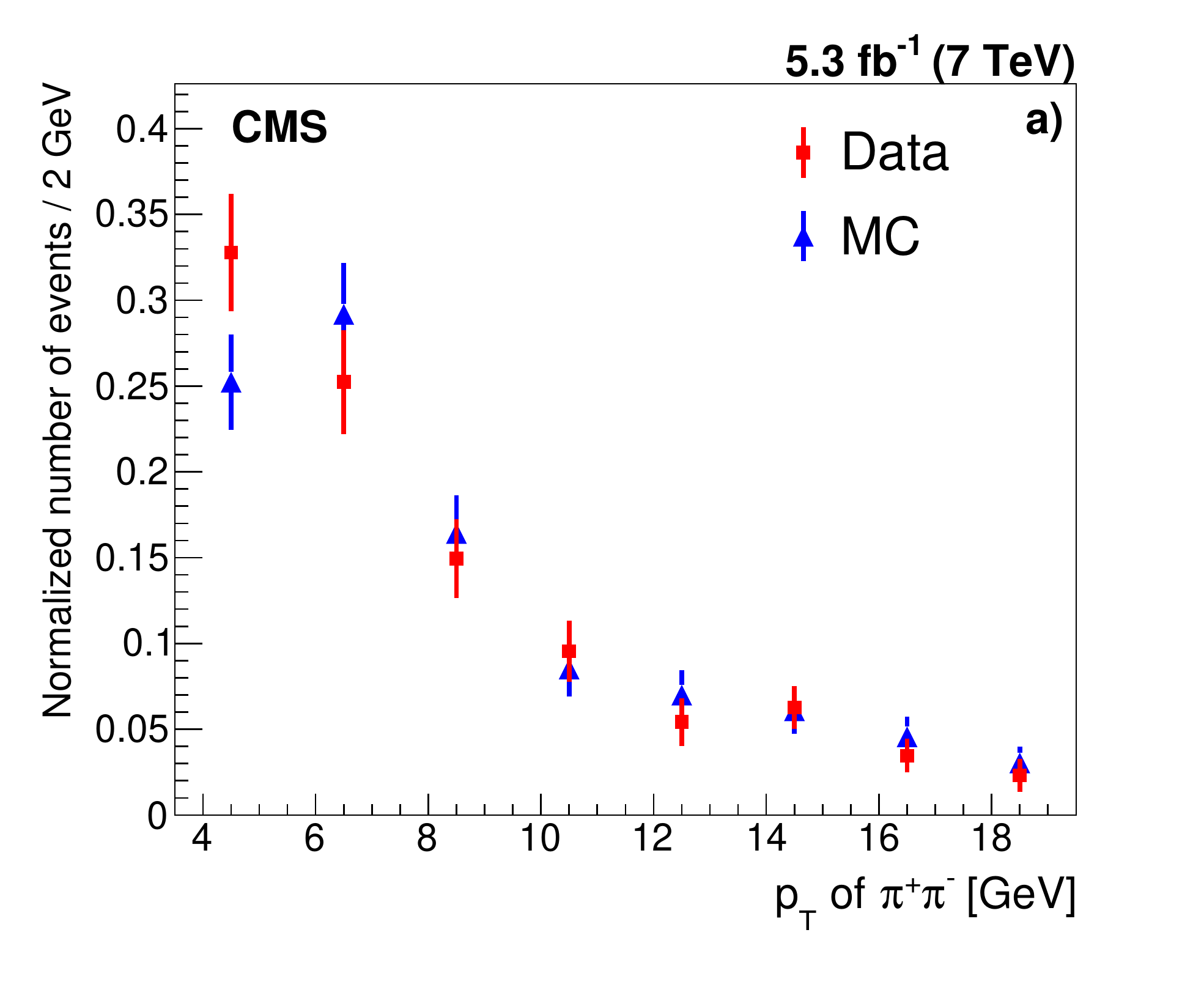}
\includegraphics[width=0.49\textwidth]{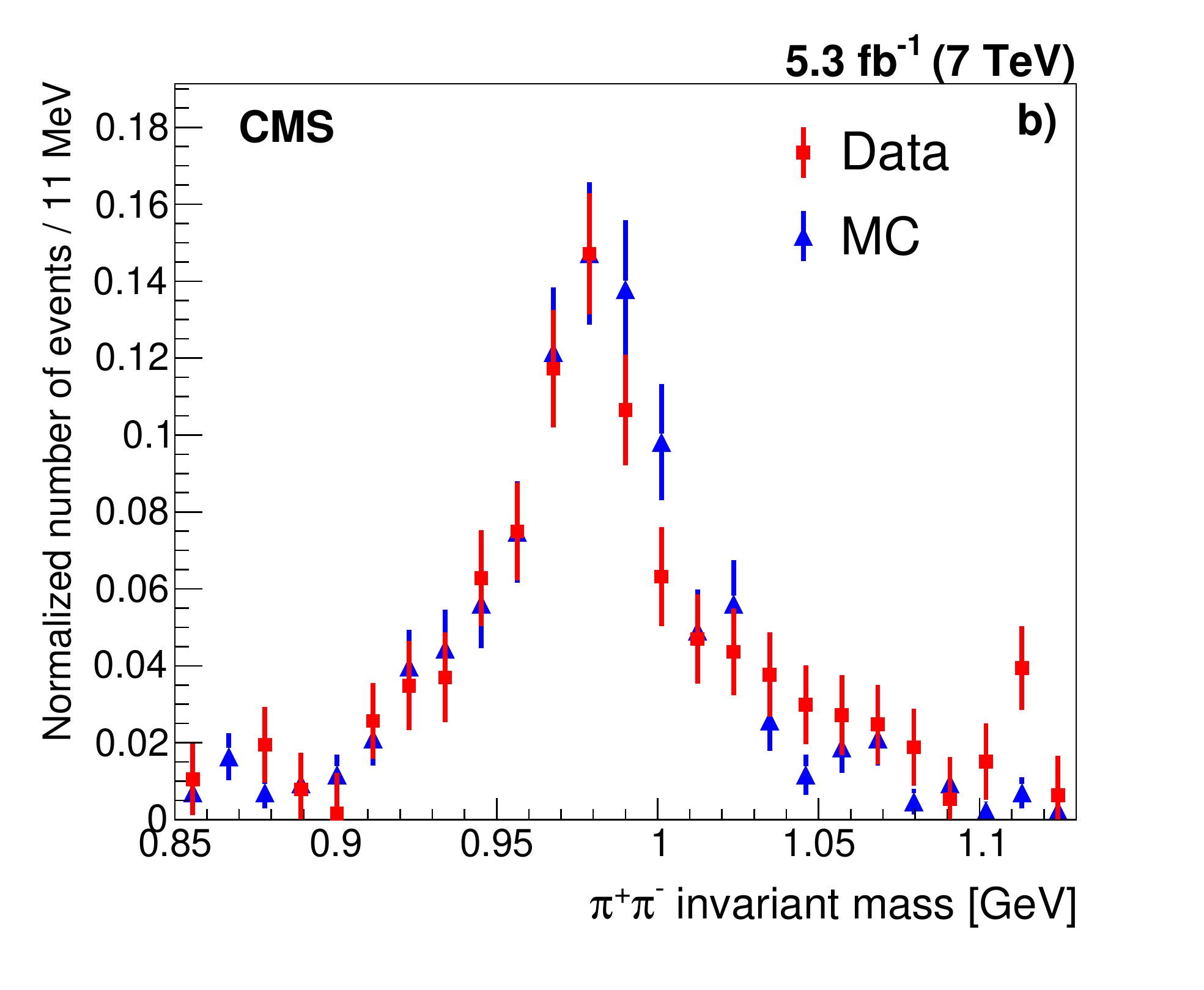}
 \caption{Comparison of normalized MC simulation (triangles) and background-subtracted data (squared) for (a) the \pt and (b)
 invariant mass distributions of the $\cPfz(\Pgpp\Pgpm)$ candidates.}
\label{fig:MCvsData}
\end{figure}

Using the MC simulation, the detection efficiencies for the two processes are calculated as the ratio of the reconstructed  and
generated yields. The $\PBzs$ meson production is simulated using \PYTHIA 6.4.24~\cite{Sjostrand:2006za} and its decays
simulated with \EVTGEN~\cite{EVTGENpackage:2001}. The $\PBzs$ mass and lifetime are set to 5369.6\MeV and 438\mum in the simulation.
The decay model used for the $\PBzs\to \JPsiX \cPfz$ decay is
a phase-space model reweighted to reflect the spin-1 structure of the $\JPsi\to \Pgmp\Pgmm$ decay.
The corresponding models for the $\PBzs\to \JPsiX\phi$ decay are:  a pseudoscalar-vector-vector with CP violation
\cite{Dunietz:1991, BABAR:2001} for the  $\PBzs$ decay, with parameters \cite{EVTGENpackage:2001} $||A_{||}||^2 = 0.24$, $||A_{0}||^2 = 0.6$,
$||A_{\perp}||^2 = 0.16$, $\phi_{||} = 2.5$, $\phi_{0} = 0$, and $\phi_{\perp} = -0.17$;
a vector-lepton-lepton model with radiation (\PHOTOS)~\cite{Golonka:2006} for the $\JPsi \to \Pgmp\Pgmm$ decay; and a vector-scalar-scalar model
\cite{EVTGENpackage:2001} for the $\phi\to \PKp\PKm$ decay.
The events are processed with a  \GEANTfour-based detector simulation \cite{GEANT4} and the same reconstruction algorithms used on
data. In order to validate the MC simulation samples, relevant kinematic and geometric variables of both simulated decay
channels are compared with the data after background subtraction and found to be in agreement. For example,
Fig. \ref{fig:MCvsData} compares the $\pt$ and invariant mass distributions of the $\cPfz(\Pgpp\Pgpm)$ candidates for
background-subtracted data and MC simulation. The $\cPfz$ width was set to 50\MeV in the MC simulation. This is consistent
with what is observed in our data as shown in the Fig. \ref{fig:MCvsData}.b.
The ratio of the detection efficiencies for the two $\PBzs$ decays is calculated to be
$\epsilon^{\phi/\cPfz}_{\text{reco}}= 1.344 \pm 0.095$, where the uncertainty reflects the limited size of simuated samples.
Using the corresponding values of $N_{\text{obs}}^{\cPfz}$, $N_{\text{obs}}^{\phi}$, and $\epsilon^{\phi/\cPfz}_{\text{reco}}$
in Eq.~(\ref{eq:BR}), we measure $R_{\cPfz/\phi}= 0.140 \pm 0.008$, where the uncertainty is statistical only.

The stability of the $R_{\cPfz/\phi}$ measurement is verified with control checks using different run periods,
selection criteria, and geometric acceptances. To study possible effects from varying run conditions, the value of
$R_{\cPfz/\phi}$ is determined for two subsamples,
found by dividing the data into two. The ratio is also measured after changing the selection criteria for the proper decay length
and \pt of the $\PBzs$ candidates and the \pt of the leading and subleading pion candidates, and by using different
azimuthal angle and $\eta$ requirements for the muons. None of these cross-checks revealed any statistically significant
bias.

\section{Systematic uncertainties}

Potential systematic uncertainties in the measurement of $R_{\cPfz/\phi}$ come from sources such as
the $\PBzs$ signal yield extraction procedure,  the relative efficiency estimation, and possible contributions
to the $\PBzs$ yields from other decays producing the $\JPsi\Pgpp\Pgpm$ and $\JPsi \PKp\PKm$ final states.

Systematic uncertainties in the signal yield extraction are estimated by changing the modeling of the signal and the background
invariant mass distributions in the likelihood fits. For the case of the $\PBzs\to\JPsi\cPfz$ mass distribution the signal shape is
changed to a double-Gaussian function and the background to an exponential function, while for the $\PBzs\to\JPsi\phi$ mass distribution
the signal is changed to a Gaussian function and its background is modelled as
a first-order polynomial function. These changes lead to a maximum variation of 2.1\% in $R_{\cPfz/\phi}$.

There are several factors that may affect the estimate of $\epsilon_{\text{reco}}^{\phi/\cPfz}$.
While the MC simulation package uses a Breit--Wigner model to simulate the $\cPfz\to \Pgpp\pi^{-}$ process,
it has been pointed out \cite{JpsiPiPiLHCb:2012, LHCbCPcomp:2014} that a Flatt\'{e} model is a better description of this decay.
To estimate the effect of the simulation model,  the Breit--Wigner model used in the simulation is compared to a Flatt\'{e} model 
in the selected $M_{\cPpipi}$ region. The difference in the models reflects a systematic error
of 5.8\% in $\epsilon_{\text{reco}}^{\phi/\cPfz}$. This is quoted as a systematic uncertainty. 
In addition, in the MC simulation the $\cPfz$ width is set to $50\MeV$. This value is varied by $\pm$10\MeV,
resulting in a systematic uncertainty of 8.6\% in $R_{\cPfz/\phi}$.
The models used in the MC simulation of the $B_s$ decays are set to phase-space~\cite{EVTGENpackage:2001} instead of the default decay models,
leading to a 6.2\% systematic uncertainty in $R_{\cPfz/\phi}$.
Finally, the statistical uncertainty in $\epsilon_{\text{reco}}^{\phi/\cPfz}$ owing to the finite number of MC events,
which corresponds to 7.1\%, is added as a systematic uncertainty.

As mentioned in the introduction, detailed studies of the $\pi^+\pi^-$ mass spectrum of the $\PBzs \to \JPsiX\pi^+\pi^-$
decay \cite{JpsiPiPiLHCb:2012, LHCbCPcomp:2014} in a mass window of 0.3--2.5\GeV,
reveal this final state to have contributions from several resonances in $M_{\cPpipi}$, and the $\cPfz$ component to range from 65.0 to 94.5\% in the
entire mass window studied by LHCb. To study the effects of the interferences and the $\cPfz$ fraction observed by LHCb in the estimate of
$\epsilon_{\text{reco}}^{\phi/\cPfz}$, the model reported in \cite{LHCbCPcomp:2014} for the lowest $\cPfz$ fraction and largest non-resonant component was
compared to the single Breit--Wigner model used in the MC simulation of the $\cPfz\to \Pgpp\Pgp$ decay.
The comparison in the selected $M_{\cPpipi}$ region
shows a variation of 5.6\% in $R_{\cPfz/\phi}$. This is quoted as a systematic uncertainty coming from this source.
It can be observed in the same LHCb study that the contaminations from other resonances in the mass region $\abs{M_{\cPpipi} - 974\MeV} < 50\MeV$ are several orders of magnitude lower
than the $\cPfz$ component, including the non-resonant S-wave.
To estimate the variation in the $\PBzs$ yield coming from these possible contributions,
the $\cPfz$ mass window is widened from 50 to 100\MeV around the $\cPfz$ mass, resulting in a variation in $R_{\cPfz/\phi}$  of 6.4\% that is quoted
as a systematic uncertainty. For the $\PBzs \to\JPsiX \PKp\PKm$ decay channel, the contribution of the S-wave in a $\phi$ mass window similar
to what is used in this analysis has been found to be negligible \cite{SwaveCDF:2010}.

Combining these uncertainties in quadrature leads to a total systematic uncertainty of 16.5\%.

\section{Summary}

Using data collected by the CMS experiment in proton-proton collisions at $\sqrt{s}=7\TeV$,
corres\-pon\-ding to an integrated luminosity of 5.3\fbinv,  $873\pm 49$ events of  $\PBzs \to\JPsi(\Pgmp\Pgmm)
\cPfz(\Pgpp\Pgpm)$ and $8377\pm 107$ events of  $\PBzs \to \JPsi(\Pgmp\Pgmm)\phi(\PKp\PKm)$
are observed. The $\cPfz$ and $\phi$ are identified in the mass ranges $\abs{M_{\cPpipi}-974\MeV}<50\MeV$ and $\abs{M_{\cPKK}-1020\MeV}<10\MeV$, respectively.
The ratio of the branching fraction of $\PBzs\to\JPsi(\Pgmp\Pgmm) \cPfz(\Pgpp\Pgpm)$ to the branching fraction of
$\PBzs \to \JPsi(\Pgmp\Pgmm)\phi(\PKp\PKm)$, $R_{\cPfz/\phi}$, is found to be
\ifthenelse{\boolean{cms@external}}{
\begin{multline}
\frac{\mathcal{B}(\PBzs\to  \JPsiX\cPfz)\,\mathcal{B}(\cPfz\to \Pgpp\Pgpm)}{\mathcal{B}(\PBzs\to  \JPsiX\phi )\,\mathcal{B}(\phi\to \PKp\PKm)}
= \\
0.140 \pm  0.008\stat \pm 0.023\syst.
\end{multline}
}{
\begin{linenomath}
\begin{equation}
\frac{\mathcal{B}(\PBzs\to  \JPsiX\cPfz)\,\mathcal{B}(\cPfz\to \Pgpp\Pgpm)}{\mathcal{B}(\PBzs\to  \JPsiX\phi )\,\mathcal{B}(\phi\to \PKp\PKm)}
= 0.140 \pm  0.008\stat \pm 0.023\syst.
\end{equation}
\end{linenomath}
}

This result is consistent with the theoretical prediction of about 0.2 \cite{Stone/Zhang:2009} and with previous measurements in different ranges of $M_{\cPpipi}$
\cite{JpsiPiPiLHCb:2012,CDFBRf0:2011,D0BRf0:2012}.

\begin{acknowledgments}
We congratulate our colleagues in the CERN accelerator departments for the excellent performance of the LHC and thank the technical and administrative staffs at CERN and at other CMS institutes for their contributions to the success of the CMS effort. In addition, we gratefully acknowledge the computing centres and personnel of the Worldwide LHC Computing Grid for delivering so effectively the computing infrastructure essential to our analyses. Finally, we acknowledge the enduring support for the construction and operation of the LHC and the CMS detector provided by the following funding agencies: BMWFW and FWF (Austria); FNRS and FWO (Belgium); CNPq, CAPES, FAPERJ, and FAPESP (Brazil); MES (Bulgaria); CERN; CAS, MoST, and NSFC (China); COLCIENCIAS (Colombia); MSES and CSF (Croatia); RPF (Cyprus); MoER, ERC IUT and ERDF (Estonia); Academy of Finland, MEC, and HIP (Finland); CEA and CNRS/IN2P3 (France); BMBF, DFG, and HGF (Germany); GSRT (Greece); OTKA and NIH (Hungary); DAE and DST (India); IPM (Iran); SFI (Ireland); INFN (Italy); MSIP and NRF (Republic of Korea); LAS (Lithuania); MOE and UM (Malaysia); CINVESTAV, CONACYT, SEP, and UASLP-FAI (Mexico); MBIE (New Zealand); PAEC (Pakistan); MSHE and NSC (Poland); FCT (Portugal); JINR (Dubna); MON, RosAtom, RAS and RFBR (Russia); MESTD (Serbia); SEIDI and CPAN (Spain); Swiss Funding Agencies (Switzerland); MST (Taipei); ThEPCenter, IPST, STAR and NSTDA (Thailand); TUBITAK and TAEK (Turkey); NASU and SFFR (Ukraine); STFC (United Kingdom); DOE and NSF (USA).

Individuals have received support from the Marie-Curie programme and the European Research Council and EPLANET (European Union); the Leventis Foundation; the A. P. Sloan Foundation; the Alexander von Humboldt Foundation; the Belgian Federal Science Policy Office; the Fonds pour la Formation \`a la Recherche dans l'Industrie et dans l'Agriculture (FRIA-Belgium); the Agentschap voor Innovatie door Wetenschap en Technologie (IWT-Belgium); the Ministry of Education, Youth and Sports (MEYS) of the Czech Republic; the Council of Science and Industrial Research, India; the HOMING PLUS programme of Foundation for Polish Science, cofinanced from European Union, Regional Development Fund; the Compagnia di San Paolo (Torino); the Consorzio per la Fisica (Trieste); MIUR project 20108T4XTM (Italy); the Thalis and Aristeia programmes cofinanced by EU-ESF and the Greek NSRF; and the National Priorities Research Program by Qatar National Research Fund.
\end{acknowledgments}
\bibliography{auto_generated}
\cleardoublepage \appendix\section{The CMS Collaboration \label{app:collab}}\begin{sloppypar}\hyphenpenalty=5000\widowpenalty=500\clubpenalty=5000\textbf{Yerevan Physics Institute,  Yerevan,  Armenia}\\*[0pt]
V.~Khachatryan, A.M.~Sirunyan, A.~Tumasyan
\vskip\cmsinstskip
\textbf{Institut f\"{u}r Hochenergiephysik der OeAW,  Wien,  Austria}\\*[0pt]
W.~Adam, T.~Bergauer, M.~Dragicevic, J.~Er\"{o}, M.~Friedl, R.~Fr\"{u}hwirth\cmsAuthorMark{1}, V.M.~Ghete, C.~Hartl, N.~H\"{o}rmann, J.~Hrubec, M.~Jeitler\cmsAuthorMark{1}, W.~Kiesenhofer, V.~Kn\"{u}nz, M.~Krammer\cmsAuthorMark{1}, I.~Kr\"{a}tschmer, D.~Liko, I.~Mikulec, D.~Rabady\cmsAuthorMark{2}, B.~Rahbaran, H.~Rohringer, R.~Sch\"{o}fbeck, J.~Strauss, W.~Treberer-Treberspurg, W.~Waltenberger, C.-E.~Wulz\cmsAuthorMark{1}
% ...
\vskip\cmsinstskip
% ...
1:~~Also at Vienna University of Technology, Vienna, Austria\\
2:~~Also at CERN, European Organization for Nuclear Research, Geneva, Switzerland\\
% ...

\end{sloppypar}
\end{document}